\documentclass{PoS}

\usepackage{amsmath}
\usepackage{upgreek}
\usepackage{grffile}

\newcommand\su[1]{\mathrm{SU}(#1)}
\newcommand\Nf{N_{\textnormal{f}}}
\newcommand\mf{m_{\textnormal{f}}}

\newcommand\chit {\chi_{\textnormal{top}}}
\newcommand\chitop\chit

\newcommand \dee {\mathrm{d}}
\newcommand \Dee {\mathrm{D}}

\def\figsubcap#1{\par\noindent\centering\footnotesize#1}

\title{
\vspace*{-3cm}
{\normalsize {\rm \hfill\textnormal{LLNL-PROC-679800} }} \\
\vspace*{2cm}
Topological observables in many-flavour QCD}

\ShortTitle{Topological observables in many-flavour QCD}

\author{Yasumichi Aoki$^a$, Tatsumi Aoyama$^a$, \speaker{Ed Bennett}$^{b\dagger}$, Masafumi Kurachi$^c$, Toshihide Maskawa$^a$, Kohtaroh Miura$^{ad}$, Kei-ichi Nagai$^a$, Hiroshi Ohki$^e$, Enrico Rinaldi$^f$, Akihiro Shibata$^g$, Koichi Yamawaki$^a$, Takeshi Yamazaki$^{h}$ (LatKMI Collaboration)\\
        $^a$ Nagoya University, Furo, Chikusa, Nagoya, Aichi, 464-8602 Japan\\
        $^b$ Swansea University, Singleton Park, Swansea SA2 8PP UK\\
        $^c$ KEK, Tsukuba, Ibaraki 305-0801 Japan\\
        $^d$ CPT, Campus de Luminy, Case 907, 163 Avenue de Luminy, 13288 Marseille cedex 9, France\\
        $^e$ RIKEN BNL Research Center, Brookhaven National Laboratory, Upton, NY 11973-5000 USA\\
        $^f$ Lawrence Livermore National Laboratory, 7000 East  Ave., Livermore, CA 94550-9234 USA\\
        $^g$ Computing Research Center, High Energy Accelerator Research Organization (KEK), Tsukuba, Ibaraki 305-0801 Japan\\
        $^h$ Graduate School of Pure and Applied Sciences, University of Tsukuba, Tsukuba, Ibaraki 305-8571, Japan\\
        \null\\
        $^\dagger$ E-mail: \email{e.j.bennett@swansea.ac.uk}
}

\abstract{$\su{3}$ gauge theory with eight massless flavours is believed to be walking, while the corresponding twelve- and four-flavour appear IR-conformal and confining respectively. Looking at the simulations performed by the LatKMI collaboration of these theories, we use the topological susceptibility as an additional probe of the IR dynamics. By drawing a comparison with $\su{3}$ pure gauge theory, we see a dynamical quenching effect emerge at larger number of flavours, which is suggestive of emerging near-conformal and conformal behaviour.}

\FullConference{The 33rd International Symposium on Lattice Field Theory\\
		 14 - 18 July  2015\\
		 Kobe International Conference Center, Kobe, Japan}

\begin{document}

\section{Introduction}
For a number of years, the LatKMI Collaboration has performed an ongoing investigation\cite{Aoki:2012eq,Aoki:2013xza} of QCD with 4, 8, 12, and 16 light flavours on the lattice, with the aim to classify them as conformal, near-conformal, or confining and chirally broken, and calculate quantities of phenomenological relevance. In particular, $\Nf=8$ QCD is a candidate for a Walking Technicolor theory, with the potential to provide a mechanism for Electroweak Symmetry Breaking and at the same time produce a (composite) Higgs.

In lattice gauge theories near the continuum limit, it is possible for the topological charge $Q$ of a computation to become frozen, causing the simulation to become non-ergodic. Since observables of physical interest are related to topological effects, and thus depend on the topological charge, it is thus necessary to compute the topological charge in order to verify whether or not our computations have sufficient ergodicity. 

Having calculated these values, we would like to use them for more than just a check of ergodicity. In particular, it would be useful if they could contribute to our ongoing work to study the infrared regime of these theories, specifically whether they are conformal or confining and chirally broken, and if the latter, whether they are QCD-like or exhibit walking-type dynamics. Bennett and Lucini have previously performed a study of this kind\cite{Bennett:2012ch} for $\su{2}$ with $\Nf=2$ adjoint flavours, and we adopt a similar procedure here. In the next section we will discuss the methodology we use for this; we will then outline the methods we have used to calculate the quantities discussed, and present preliminary results for the theories mentioned above.

\section{(Near-)conformal topology}
An IR-conformal theory becomes confining when deformed by a fermion mass. In the process, a scale is added (the fermion mass) which is intrinsically heavy---thus when considering gluonic IR observables, the theory should be indistinguishable from the equivalent quenched theory. One such observable is the topological susceptibility $\chitop$, which we calculate as 
\begin{equation}
	\chitop = \frac{1}{V}\left(\langle Q^2 \rangle - \langle Q \rangle^2\right)\label{eq:suscept}
\end{equation}
where $V$ is the lattice volume, and $\langle Q \rangle$ is expected to be zero. Since in IR-conformal theories, the deforming mass induces non-trivial RG corrections to the lattice spacing, we must consider only dimensionless ratios, and since we will compare with results from a theory without fermions, we consider primarily pure gluonic observables to scale the results.

Additional information may be gained by probing the dependence of $\chitop$ on the fermion mass directly---specifically, checking whether the $\chitop$ obeys the predictions of either chiral perturbation theory ($\upchi$PT) or conformal hyperscaling. For the former case we fit
\begin{align}
	\chitop &= C\mf + f(a) & \Rightarrow \chitop^{1/4} &= (C\mf + f(a))^{1/4}\;;
\end{align}
while for the latter we use
\begin{equation}
	\chitop^{1/4} = A\mf^{1 / (1+\gamma_*)}\;.
\end{equation}
Here, $C$, $A$, and $f(a)$ are constants to be determined, and $\gamma_*$ is the anomalous dimension of the chiral condensate.

\section{Computations}
The topological charge is defined in the continuum as
\[
        Q = \int \dee^4x q(x)\;,\quad q(x)=\frac{1}{32\pi^2}\epsilon_{\mu\nu\rho\sigma}F_{\mu\nu}F_{\rho\sigma}\;.
\]
To find the lattice equivalent, we replace the integral by a sum, and obtain the equivalent of the field strength $F_{\mu\nu}$ by taking the path-ordered product of link variables around a clover-shaped path. We encounter a problem however when we apply this to gauge configurations as produced by a typical Monte Carlo process: namely that ultraviolet fluctuations dominate over the topological contribution to the charge. In principle these cancel out across the lattice volume, but in practise the topological contribution is vastly smaller than the precision error of the UV fluctuations, and so the signal is lost.

We therefore need to suppress these contributions. We use the gradient flow, as suggested by L\"uscher \cite{Luscher:2010iy}, for this purpose. The gradient flow defines a flowed field $B_\mu(t,x)$ at flow time $t$ as
\begin{align*}
        \frac{\dee}{\dee t}B_\mu &= \Dee_\nu G_{\nu\mu}\, & \left. B_\mu \right|_{t=0} &= A_\mu\;,\\
        G_{\mu\nu} &= \partial_\mu B_\nu - \partial_\nu B_\mu + [B_\mu, B_\nu]\;, & \Dee_\mu &= \partial_\mu + \left[B_\mu, \frac{\dee}{\dee t}\right]\;,
\end{align*}
where the flow starts at $B_\mu(0,x) = A_\mu(x)$, the physical gauge field. Integrating numerically from $A_\mu$ allows calculation of $B_\mu(t,x)$ at arbitrary $t$. We do this using the Runge--Kutta-like scheme also outlined by L\"uscher \cite{Luscher:2010iy}. The flow has a smearing effect on the gauge configuration, with a characteristic radius dependent on the flow time as $r=\sqrt{8t}$; it is this effect that allows extracting the topological charge.

As mentioned above, in some cases close to the continuum and chiral limits, the topology may tend to freeze at a single or a very restricted range of topological charge. This will render equation \eqref{eq:suscept} for the topological susceptibility invalid, as both $\langle Q\rangle$ and $\langle Q^2\rangle$ will be distorted by the insufficient sampling. In these cases we extract further information from the topological charge density distribution by considering subvolumes of the full lattice rather than the whole lattice volume \cite{Brower:2014bqa}. This allows the excluded volume to act as a reservoir of topological charge, to and from which instantons may move.

Further information may be extracted by probing the distribution of topological charge: peaks and troughs in this distribution correspond to instantons and anti-instantons respectively. The height of the peak (trough) $q_{\textnormal{peak}}$ is directly related to the characteristic size $\rho$ of the (anti)instanton, as \cite{Bennett:2012ch}
\[
	q_{\textnormal{peak}} = \frac{6}{\pi^2 \rho^4}\;.
\]

Finally, the gradient flow, having been used to access the topological charge, may also be used at no extra cost to extract a scale\cite{Luscher:2010iy} $t_0$, defined as the value of the flow time $t$ for which $t^2\langle E\rangle = 0.3$, where $E$ is a lattice discretized version of the continuum relation $E=\frac{1}{4}G_{\mu\nu}G_{\mu\nu}$. $G_{\mu\nu}$ may be calculated in two ways: via the average plaquette, or via a four-plaquette clover operator.

\section{Results}
\begin{figure}
\begin{center}

\parbox{0.49\columnwidth}{\includegraphics[width=0.49\columnwidth]{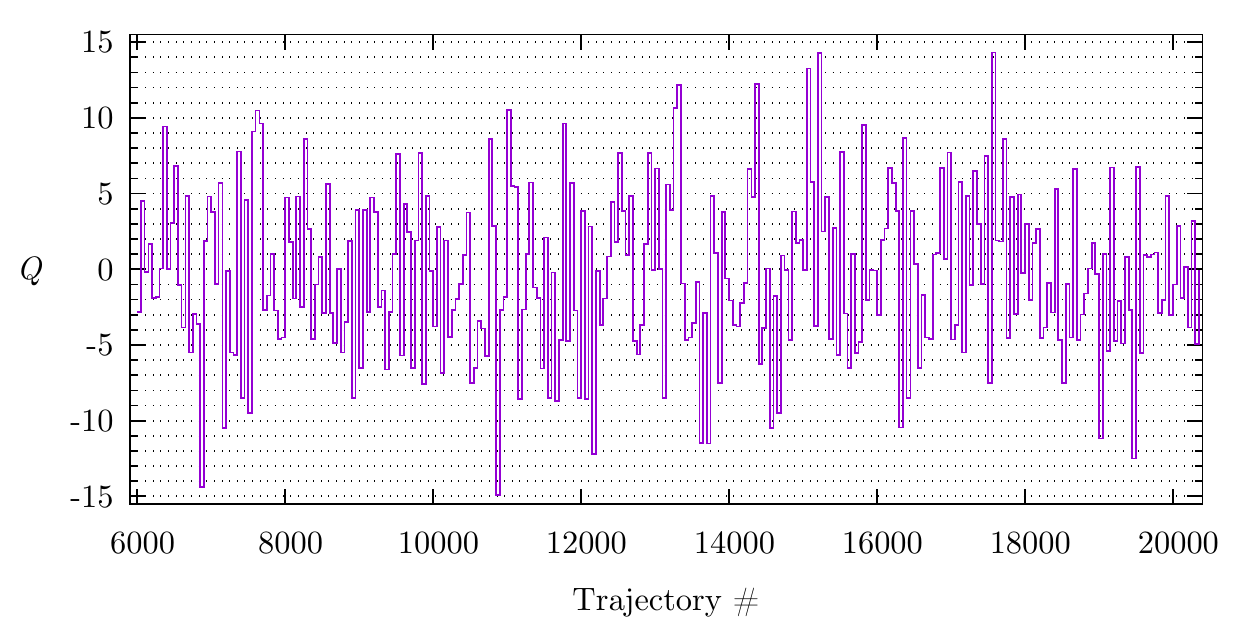}
\figsubcap{$\Nf=4,\beta=3.7,m=0.01,L=20$}}
\parbox{0.49\columnwidth}{\includegraphics[width=0.49\columnwidth]{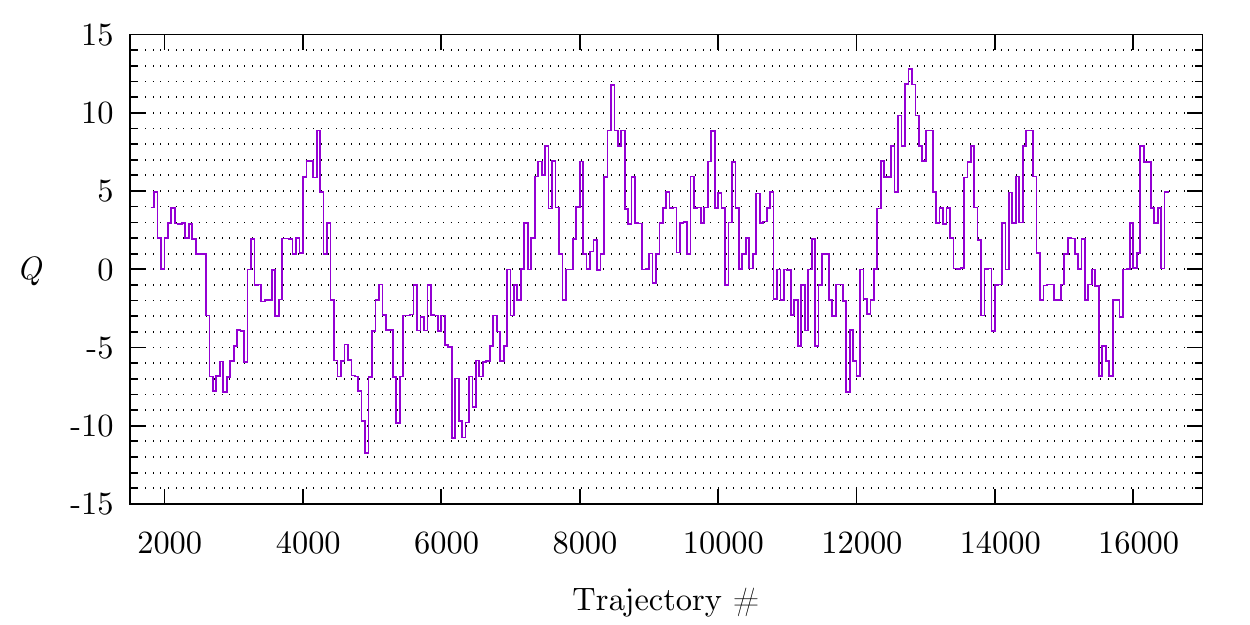}
\figsubcap{$\Nf=8,\beta=3.8,m=0.04,L=30$}}

\parbox{0.49\columnwidth}{\includegraphics[width=0.49\columnwidth]{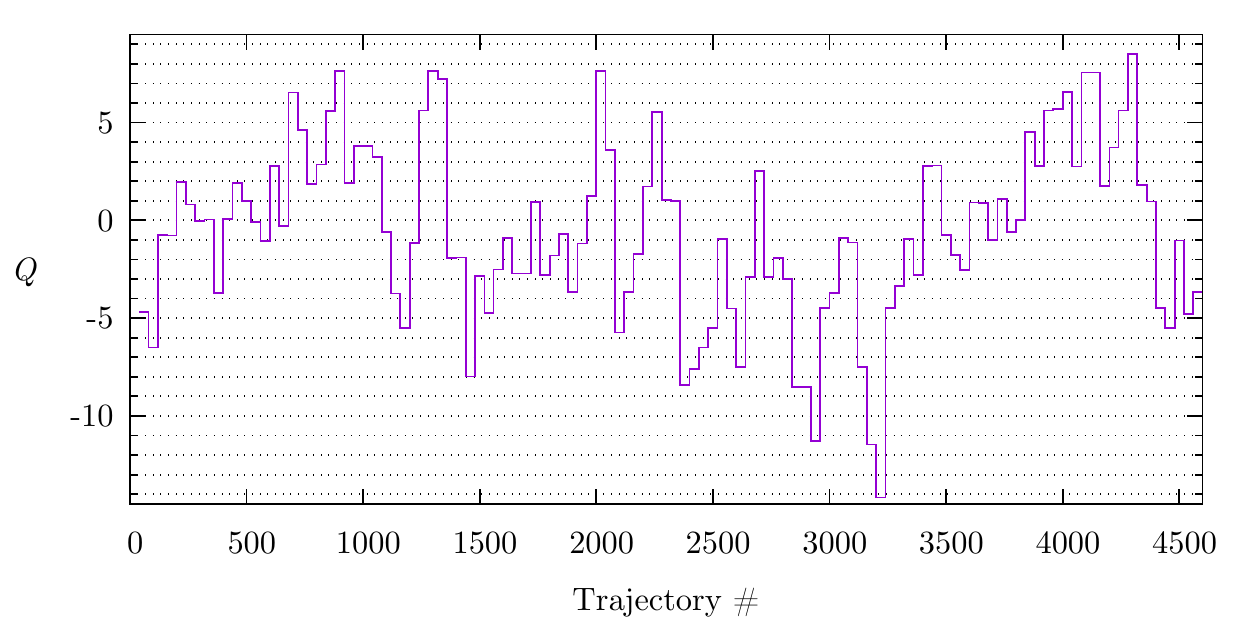}
\figsubcap{$\Nf=12,\beta=3.7,m=0.16,L=18$}}
\parbox{0.49\columnwidth}{\includegraphics[width=0.49\columnwidth]{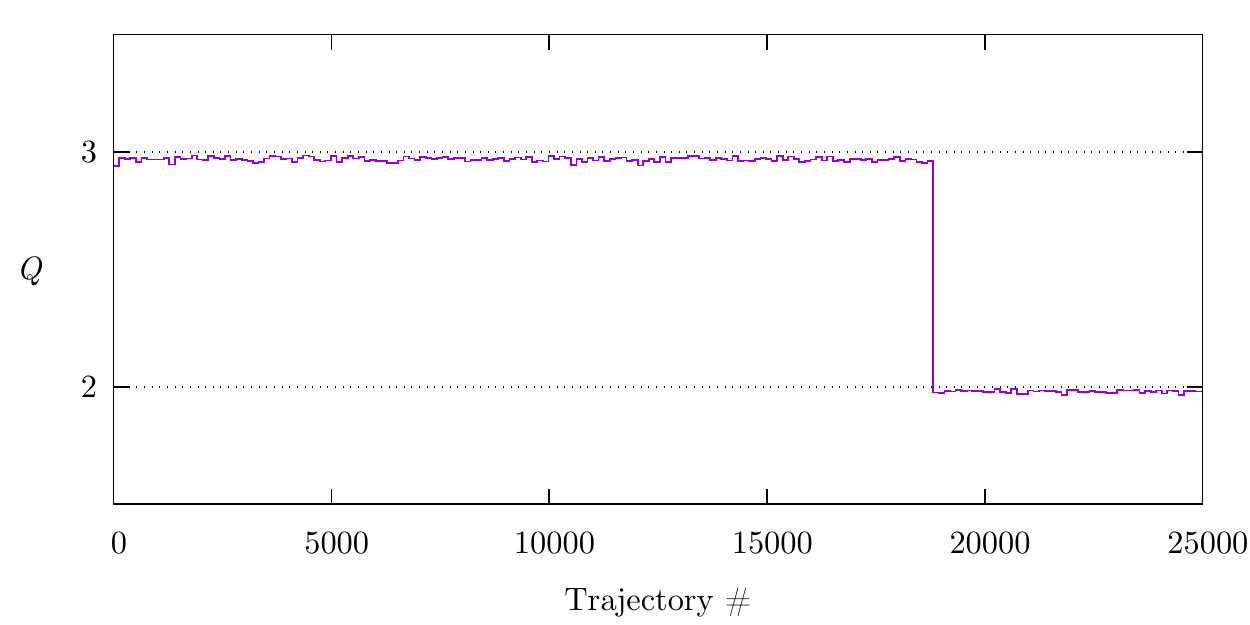}
\figsubcap{$\Nf=12,\beta=4.0,m=0.04,L=36$}}
 \caption{Sample histories of the topological charge $Q$ for the parameter sets shown. $\Nf=12,\beta=4.0$ shows the most substantial freezing observed outside of $\Nf=16$.}
\label{fig:Qhist}
\end{center}

\end{figure}

\begin{figure}
\begin{center}

\parbox{0.49\columnwidth}{\includegraphics[width=0.49\columnwidth]{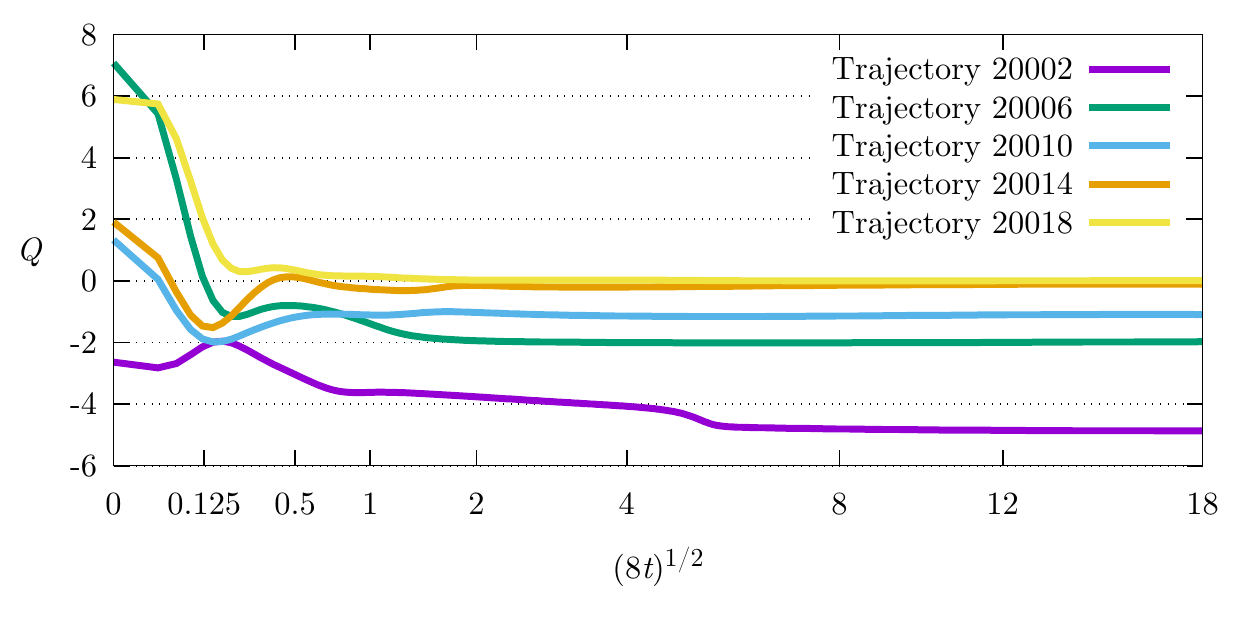}
\figsubcap{$\Nf=8,\beta=3.7,m=0.06,L=24$}}
\parbox{0.49\columnwidth}{\includegraphics[width=0.49\columnwidth]{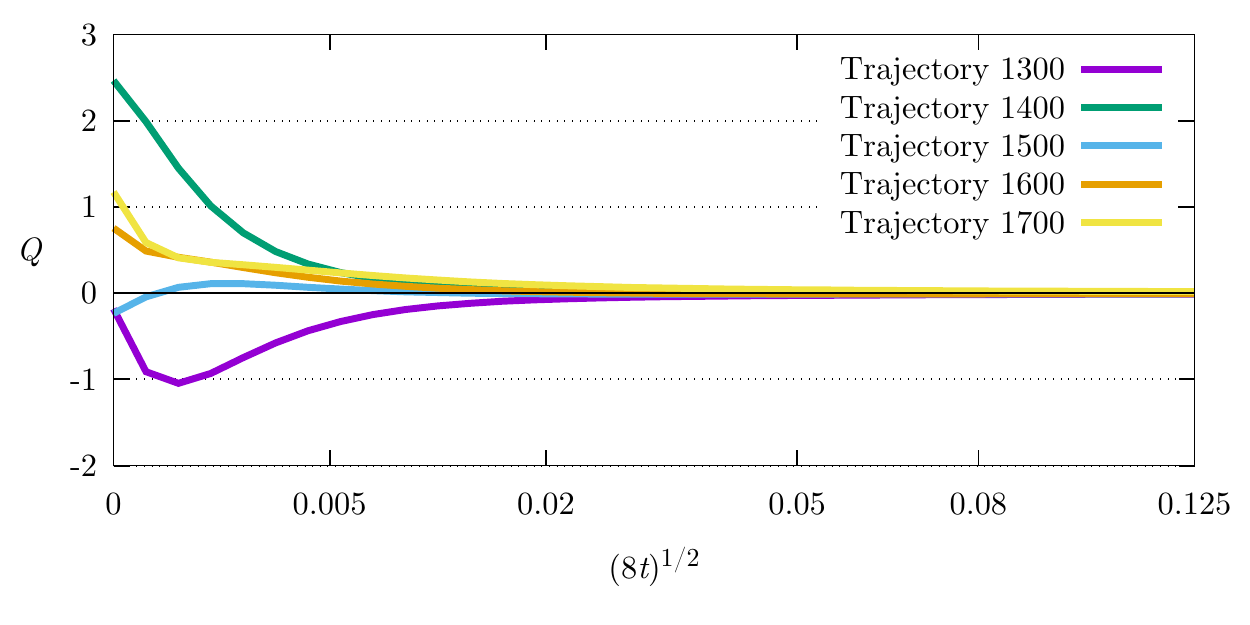}
\figsubcap{$\Nf=16,\beta=12,m=0.015,L=48$}}

\caption{History of the topological charge as a function of the gradient flow scale for selected parameter sets for $\Nf=8$ and $\Nf=16$. $\Nf=16$ is rapidly suppressed to zero, compared to $\Nf=8$ which takes longer to equilibrate and gives a variation in topological charge.}
\label{fig:flows}

\end{center}
\end{figure}

We study pre-existing $\su{3}$ gauge configurations generated by the LatKMI collaboration with one, two, three, and four staggered flavours, equivalent to $\Nf=4$ ($\beta=3.8$), $8$ ($\beta=3.8$), $12$ ($\beta=3.7,4.0$), and $16$ ($\beta=12$), with the tree-level Symanzik gauge action and the HISQ fermionic action. In addition, pure gauge $\su{3}$ configurations were generated using both the Wilson ($5.7 \le \beta \le 6.2$) and the tree-level Symanzik gauge ($4.0\le\beta\le5.0$) actions, for comparison. The string tensions in the former case were obtained from \cite{Lucini:2001ej}

As a zeroth-order step, we reviewed the topological charge histories of all sets of $(\Nf,\beta,\mf,L)$ studied. We find that $\Nf=4$ is highly ergodic, as are most parameters for $\Nf=8$, with mild loss in ergodicity emerging at the lowest values of $am$. $\Nf=12$ suffers from more ergodicity problems, with very long regions of frozen $Q$, particularly at low values of $am$. Example $Q$ histories for flavoured theories are shown in Fig.~\ref{fig:Qhist}. $\Nf=0$ (not shown) was ergodic at low beta, while becoming near-frozen at high $\beta$. In the cases where the evolution was insufficiently ergodic, the subvolume method mentioned above was employed to extract $\chitop$. Finally, the topology of $\Nf=16$ is completely suppressed, with no local minima or maxima in the topological charge density distribution indicative of the presence of a topological excitation. (An example set of cooling histories is shown in Fig.~\ref{fig:flows}.) Owing to this lack of data, $\Nf=16$ will not be discussed further. 

\begin{figure}
\begin{center}

\parbox[b][][b]{0.42\columnwidth}{\includegraphics[width=0.42\columnwidth]{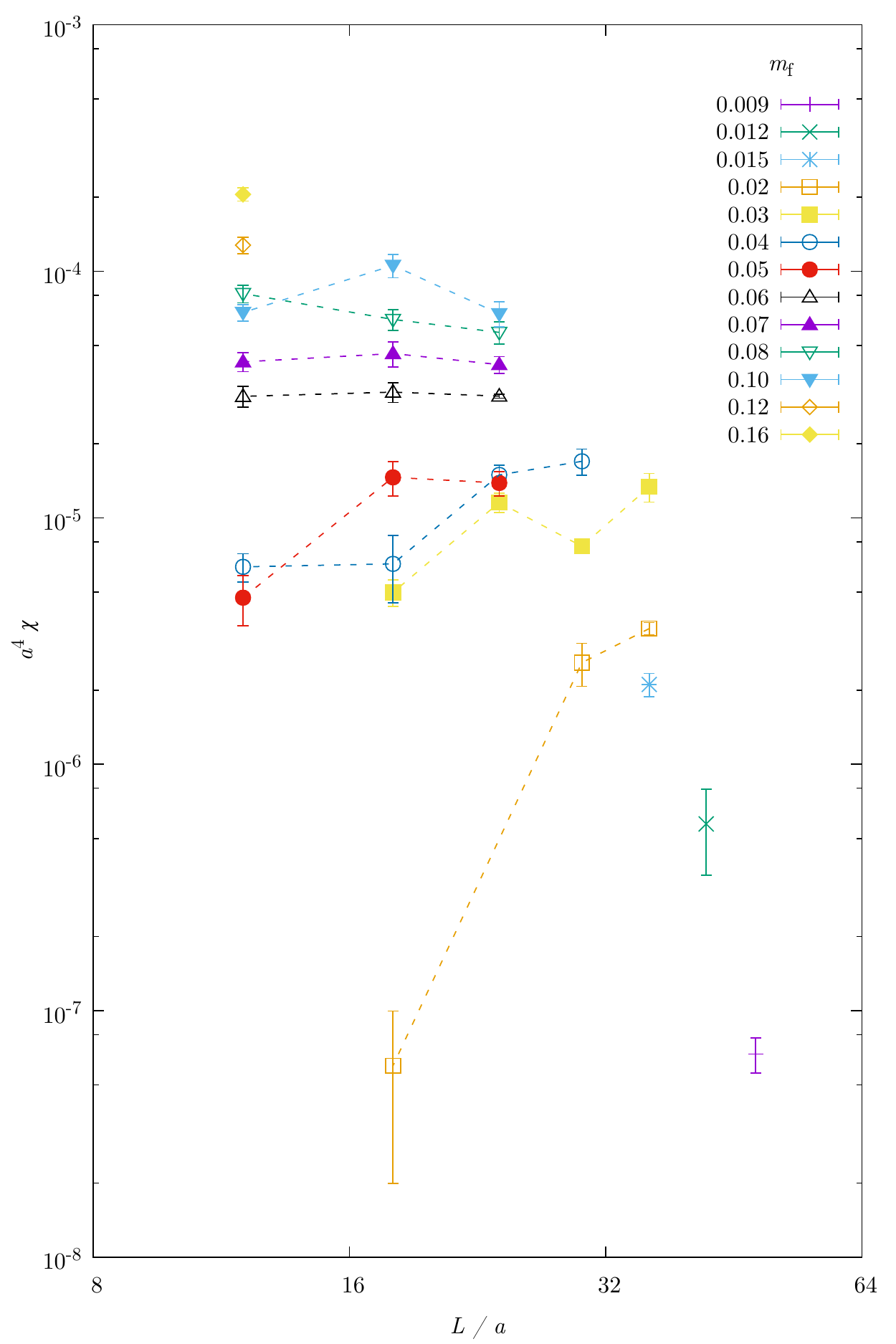}\vspace{4pt}
\figsubcap{(a)}}
\parbox[b][][b]{0.57\columnwidth}{\includegraphics[width=0.57\columnwidth]{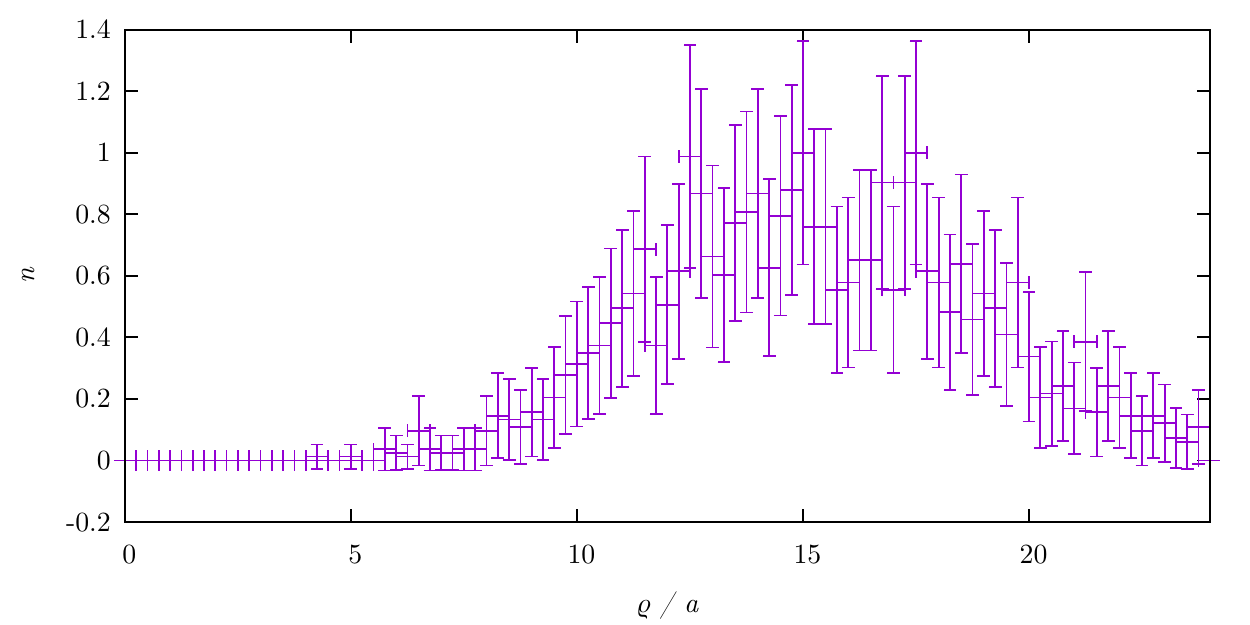}\vspace{-6pt}
\figsubcap{(b)}

\includegraphics[width=0.57\columnwidth]{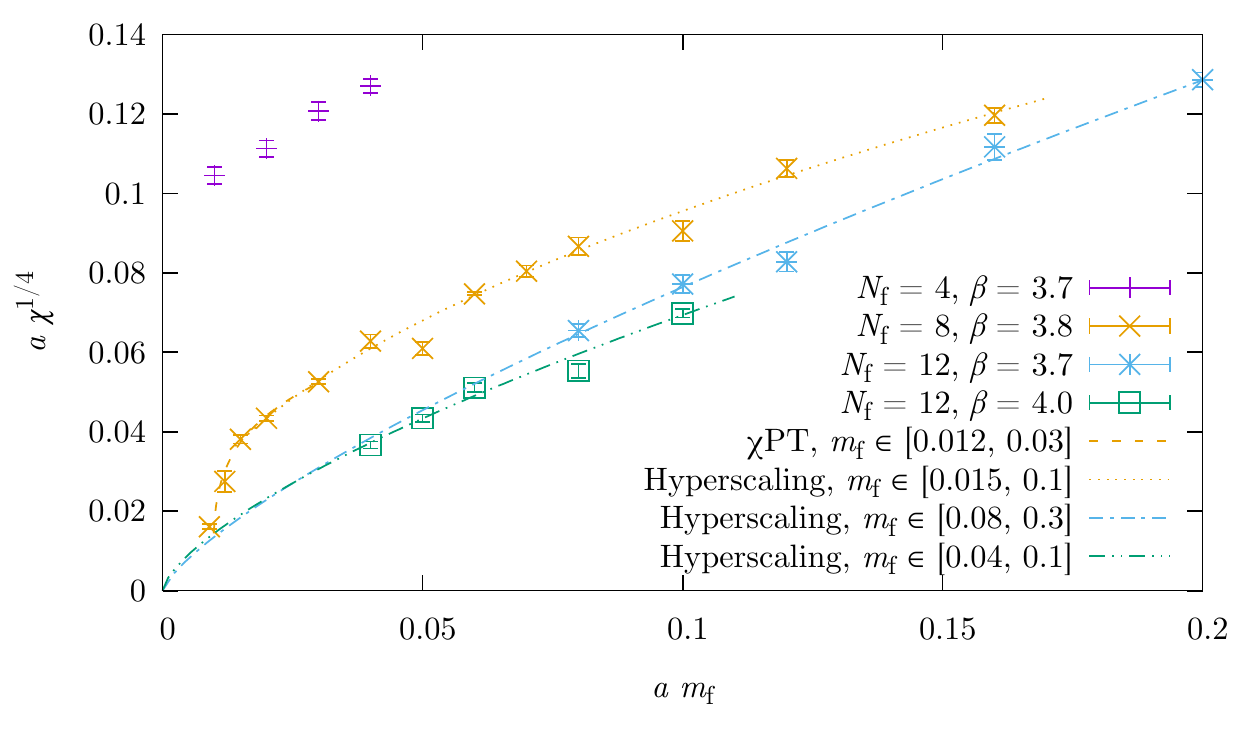}\vspace{-6pt}
\figsubcap{(c)}}

\caption{(a) Finite-volume of all available values of $\chitop$ for $\Nf=8$. (b) The instanton size distribution for $\Nf=0$, $\beta=5.0$. (c) $\chitop$ as a function of $\mf$, including conformal and $\upchi$PT fits.}
\label{fig:threeplots}

\end{center}
\end{figure}

We now move to look at the topological susceptibility. As a check that finite-volume artefacts were under control, we investigated how $\chit$ varied as a function of the spatial lattice extent $L$ for each value of $\mf$ for $\Nf=8$. The resulting plot is shown in Fig.~\ref{fig:threeplots}(a); in the majority of cases the susceptibility stabilises as $L$ increases. 

Fig.~\ref{fig:threeplots}(b) shows another indication of finite volume effects: the instanton size distribution. In the case of $\Nf=0$, $\beta=5.0$, it was found that the distribution came very close to spilling over the lattice volume---i.e. the instantons would not fit on the lattice. This ensemble was therefore not considered further; all lower values of $\beta$ had distributions that tended to zero in advance of the lattice size.

In Fig.~\ref{fig:threeplots}(c), $a\chitop^{1/4}$ for $\Nf=4$, $8$, and $12$ is plotted as a function of the fermion mass $a\mf$. We observe that for $\Nf=4$, $\chitop$ tends to a non-zero value in the limit $\mf\rightarrow0$, contrary to expectations. This is likely due to a staggered taste symmetry breaking effect at the lower mass points; if so, then it would disappear closer to the continuum limit. Work is underway to test whether this is the case; in the mean time, the $\Nf=4$ results must be regarded cautiously. 

Turning our attention to $\Nf=8$ and $12$, we fit the former with both conformal and $\upchi$SB ant\"atze, and the latter with a conformal ansatz only. For $\Nf=8$, the fits do not distinguish between conformal and confining behaviour, since both fits are equally consistent with the data; however, if we assume the conformal ansatz, then we obtain a value of $\gamma_*=1.04(5)$, consistent with estimates from spectral hyperscaling \cite{Aoki:2013xza}. For $\Nf=12$, the conformal fits are consistent with the data giving $\gamma_*=0.33(6)$ and $0.47(10)$ respectively for $\beta=3.7$ and $4.0$; these are also consistent with estimates from spectral hyperscaling.

\begin{figure}
\begin{center}

\parbox{0.49\columnwidth}{\includegraphics[width=0.49\columnwidth]{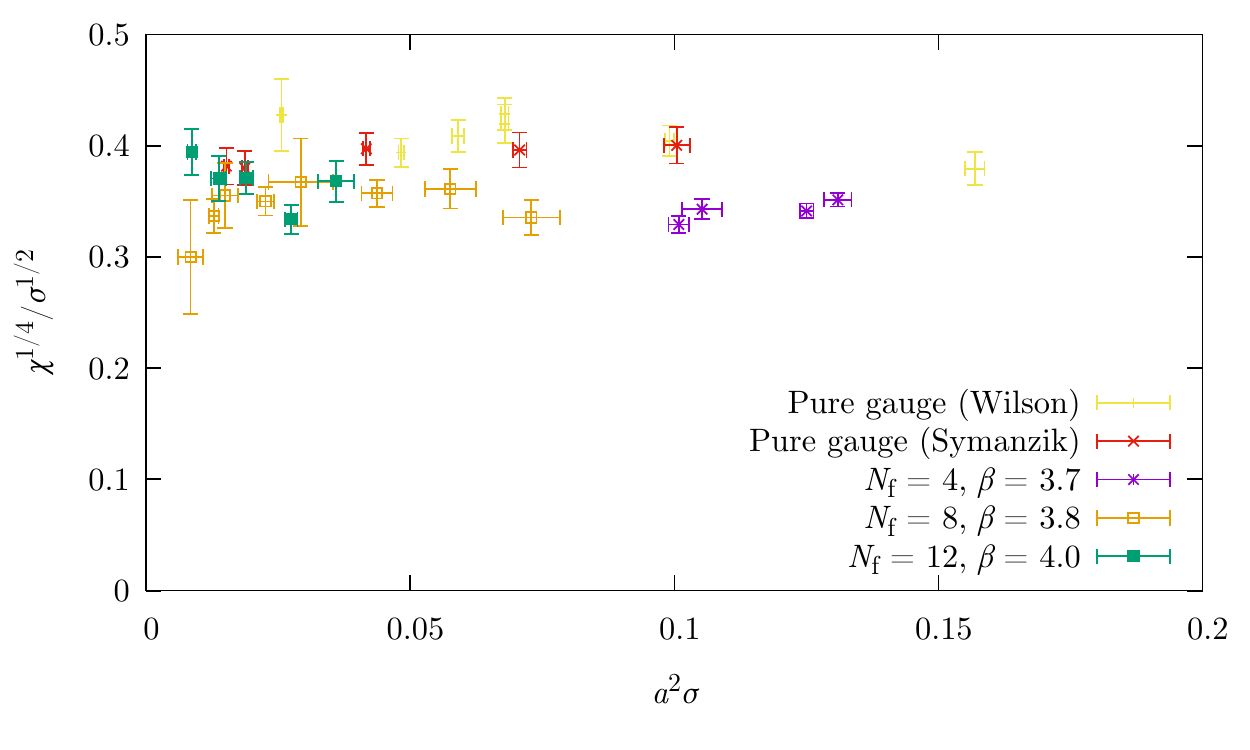}
\figsubcap{(a)}}
\parbox{0.49\columnwidth}{\includegraphics[width=0.49\columnwidth]{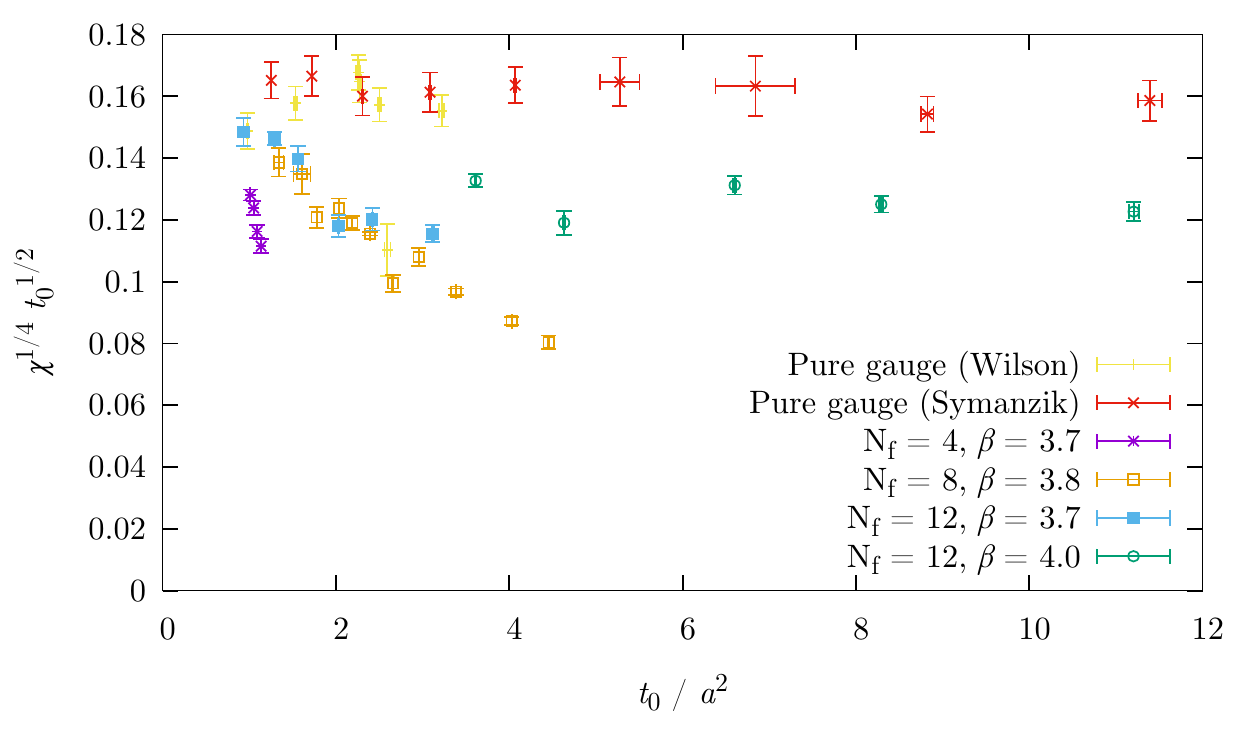}
\figsubcap{(b)}}

\vspace{-6pt}
\caption{(a) $\chitop$ scaled by the string tension $\sigma$, as a function of $a^2 \sigma$, and (b) $\chitop$ scaled by the gradient flow scale $t_0$, as a function of $t_0 / a^2$.}
\label{fig:scaling}

\end{center}
\end{figure}

In Fig.~\ref{fig:scaling}(a), $\chitop$ is plotted as a dimensionless ratio to the string tension, $\chitop^{1/4} / \sqrt{\sigma}$. The observed ratio for $\Nf=0$ is consistent with flat, with the hinted gradient depending on whether the Wilson or Symanzik-improved gauge action is used. $\Nf=4$ has a positive gradient, being close to the pure gauge theory at large $\sigma$ but diverging as $\sigma$ decreases. This is unlikely to change with the removal of the taste-symmetry breaking effects, as the positive gradient would need to become more positive to allow for zero susceptibility in the chiral limit. $\Nf=12$ is consistent with flat or a shallow negative gradient, and agrees with the pure gauge theory in the limit $\sigma\rightarrow 0$. As discussed above, this is a tentative signal of conformality.

$\Nf=8$ is consistent with $\Nf=12$ at high $\sigma$, but turns over to have a positive gradient (like $\Nf=4$) at small $\sigma$. This could be considered to be indicative of near-conformal/walking-type behaviour, with quenched behaviour at large deforming masses, turning into chirally broken behaviour as the chiral limit is approached.

We also consider the scaling of the susceptibility relative to the gradient flow scale $t_0$. This is plotted in Fig.~\ref{fig:scaling}(b): we see that in the quenched ($t_0 / a^2 \rightarrow 0$) limit, all theories become consistent, as we would expect, while the behaviour change as $t_0 / a^2$ increases. $\Nf=4$ rapidly diverges from the flat behaviour of the pure gauge theory, while $\Nf=8$ and $12$ diverge more slowly. $\Nf=12$ flattens off (emulating the flat gradient of the pure gauge theory) at higher $t_0/a^2$, a behaviour not seen in the other theories.

\section{Conclusions}
We have performed a study of the topological charge and susceptibility of QCD with 4, 8, and 12 flavours. We find that $\chitop$ for both $\Nf=8$ and $12$ may be fitted using a conformal ansatz, and gives values for the chiral condensate anomalous dimension ($\gamma_*=1.04(5)$ for $\Nf=8$, and $0.33(6)$ and $0.47(10)$ for $\Nf=12$ at $\beta=3.7$ and $4.0$ respectively) consistent with those obtained from other methods (provided the ansatz is presumed to be well-motivated for $\Nf=8$). Studying the scaling of the susceptibility with the string tension $\sigma$ and the gradient flow scale $t_0$, the $\Nf=4$ theory is qualitatively distinct from the pure gauge theory, as would be expected of a confining theory, while $\Nf=8$ and $\Nf=12$ show signs of being similar both to each other and to the pure gauge theory as we would expect for a conformal theory. However, differences between all three theories can be seen; further study is ongoing to clarify the meaning of these differences. A study of $\Nf>12$ that was not volume-constrained would provide useful insight into this, as would data for $\su{2}$ with fundamental matter.

\section{Acknowledgements}
Computations have been carried out on the $\varphi$ computer at KMI, the CX400 machine at the Information Technology Center in Nagoya University, and a workstation in Swansea University. This work is supported by the JSPS Grant-in-Aid for Scientific Research (S) No.22224003, (C) No.23540300 (K.Y.), for Young Scientists (B) No.25800139 (H.O.) and No.25800138 (T.Y.), and also by the MEXT Grants-in-Aid for Scientific Research on Innovative Areas No.23105708 (T.Y.), No. 25105011 (M.K.). E.R. acknowledges the support of the U.S. Department of Energy under Contract DE-AC52-07NA27344 (LLNL).  The work of H.O. is supported by the RIKEN Special Postdoctoral Researcher program.

\bibliographystyle{JHEP}
\bibliography{proceeding}

\end{document}